\documentstyle[twoside,fleqn,espcrc2,epsf]{article}

\def\spose#1{\hbox to 0pt{#1\hss}}
\def\ltapprox{\mathrel{\spose{\lower 3pt\hbox{$\mathchar"218$}}
 \raise 2.0pt\hbox{$\mathchar"13C$}}}
\def\gtapprox{\mathrel{\spose{\lower 3pt\hbox{$\mathchar"218$}}
 \raise 2.0pt\hbox{$\mathchar"13E$}}}
\def\inapprox{\mathrel{\spose{\lower 3pt\hbox{$\mathchar"218$}}
 \raise 2.0pt\hbox{$\mathchar"232$}}}

\newcommand{\AmS}{{\protect\the\textfont2
  A\kern-.1667em\lower.5ex\hbox{M}\kern-.125emS}}

\title{Hadron Spectroscopy with Dynamical Wilson Fermions at
$\beta=5.3$\thanks{Presented by U.~M.~Heller}}

\author{K.~M.~Bitar\address{SCRI, The Florida State University, Tallahassee,
                            FL 32306-4052, USA},
        T.~A.~DeGrand\address{Department of Physics, University of
                              Colorado, Boulder, CO 80309, USA},
        R.~Edwards$^a$,
        Steven~Gottlieb\address{Department of Physics, Indiana University,
                                Bloomington, IN 47405, USA},
        U.~M.~Heller$^a$, A.~D.~Kennedy$^a$,
        J.~B.~Kogut\address{Department of Physics, University of Illinois,
                            1110 W. Green St., Urbana, IL 61801, USA},
        A.~Krasnitz$^c$,
        W.~Liu\address{Thinking Machines Corporation, Cambridge,
                       MA 02142, USA},
        M.~C.~Ogilvie\address{Department of Physics, Washington University,
                              St.~Louis, MO 63130, USA},
        R.~L.~Renken\address{Department of Physics, University of Central
                             Florida, Orlando, FL 32816, USA},
        D.~K.~Sinclair\address{HEP Division, Argonne National Laboratory,
                               9700 S. Cass Ave., Argonne, IL 60439, USA},
        R.~L.~Sugar\address{Department of Physics, University of California,
                            Santa Barbara, CA 93106, USA},
        D.~Toussaint\address{Department of Physics, University of Arizona,
                             Tucson, AZ 85721, USA},
        and K.~C.~Wang\address{School of Physics, University of New South
                               Wales, PO Box 1, Kensington,
                               NSW 2203, Australia}}

\begin{document}

\begin{abstract}
We present results from simulations of lattice QCD using two flavors of
dynamical Wilson fermions at a lattice coupling $\beta=5.3$ on $16^3 \times
32$ lattices at two hopping parameters, $\kappa=0.1670$ and $0.1675$, leading
to $m_\pi \approx 0.44$ and $0.33$ respectively.  We show spectroscopy for
S-wave hadrons and compare our results to other recent simulations with
dynamical Wilson fermions.
\end{abstract}

\maketitle


There are two popular ways of discretizing the Dirac operator and action on
the lattice. Wilson and staggered fermions each have some advantages and
some drawbacks, and neither are entirely satisfactory. Staggered fermions
have ``good" chiral behavior with one $U(1) \times U(1)$ chiral symmetry,
protecting massless quarks. But the spin/flavor assignment is non-trivial
and really valid only in the continuum limit, and an exact simulation
algorithm requires multiples of four flavors. For Wilson fermions the
chiral symmetry is explicitly broken and its recovery requires fine tuning.
On the other hand, spin/flavor assignments are as in the continuum theory
and the exact algorithm requires only multiples of two flavors. Of course,
in the continuum limit, both formulations should lead to identical physics
results. It is, therefore, important to check whether this really holds.

To date most simulations with dynamical fermions employed staggered
fermions. This is due to the general feeling that Wilson fermions are much
more difficult to simulate \cite{doug91} and that it might be even
impossible to achieve light pions on finite lattices \cite{light}. Indeed,
in the simulations with Wilson fermions done until now the ratios $m_\pi /
m_\rho$ achieved were $\gtapprox 0.7$, while those for staggered fermions
$\gtapprox 0.5$ \cite{doug91}, with the most recent results reported at
this conference being $\approx 0.4$.

The HEMCGC (``High Energy Monte Carlo Grand Challenge") collaboration
attempted to bring dynamical Wilson fermion simulations closer to the
status of staggered fermions. Here we report on some preliminary results of
this effort \cite{GC92}.

The simulations were done on the SCRI CM-2, with a lattice size of $16^3
\times 32$, using the fast CMIS \cite{cmis} inverter, running at a
sustained speed of about 3 Gflops on half the machine. We chose a gauge
coupling $\beta = 5.3$, somewhat smaller than in typical runs with two
flavors of staggered fermions, since the renormalization of the coupling
for Wilson fermions is bigger and we did not want too small a lattice
spacing and hence too small a physical volume. We used two values of
$\kappa$, $0.1670$ and $0.1675$. For $\kappa = 0.1670$ we used a conjugate
gradient residual of $1 \times 10^{-5}$ --- we use the normalization
conventions of \cite{hemcgc90} --- and, after thermalization, time steps
$dt = 0.017$ for $425$ trajectories, $dt = 0.02$ for $1000$ trajectories
and $dt = 0.01$ for the last $100$ trajectories. These choices gave
acceptance rates of about $60\%$, $45\%$ and $80\%$ respectively. In total,
we measured propagators on $305$ configurations, separated by $5$
trajectories. For $\kappa = 0.1675$ we used a time step $dt = 0.0069$
throughout. During the warm-up we used a CG residual of $1 \times 10^{-5}$
and observed the acceptance rate dropping from about $80$ to $\sim 40\%$.
We then lowered the CG residual to $3 \times 10^{-7}$, after some tests
\cite{slgt}, after which the acceptance rate increased to about $90\%$.
Here we measured so far on $159$ configurations separated by 3 trajectories.

\begin{figure}[htb]
\epsfxsize=\columnwidth
\epsffile{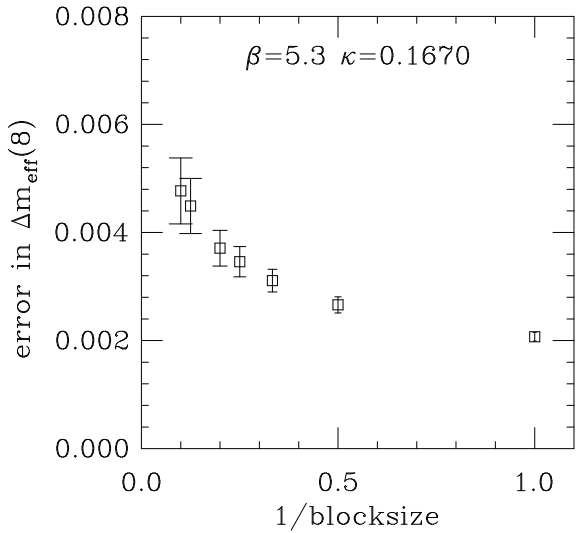}
\caption{Example of the extrapolation of errors for the effective mass at
distance 8}
\label{fig:error}
\end{figure}

Masses were extracted from fully correlated
$\chi^2$-fits of the propagators to a single exponential. Errors quoted in
Table~\ref{tab:masses} or shown in the figures were obtained from an
extrapolation in 1/blocksize from blocks of 4 and 8 measurements. An
example of the error extrapolation based upon which this choice was made is
shown in Fig.~\ref{fig:error}.

\begin{figure}[htb]
\epsfxsize=\columnwidth
\epsffile{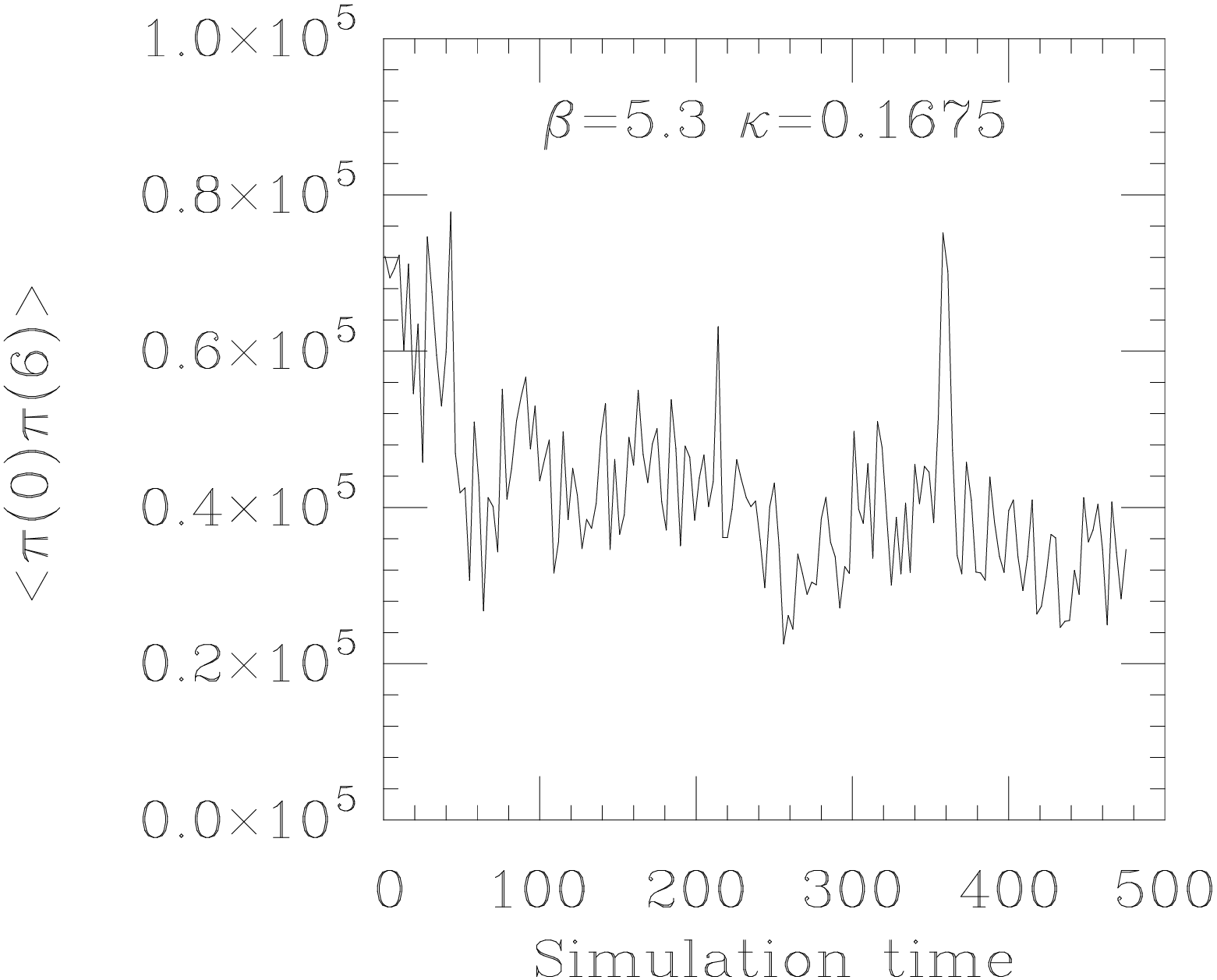}
\caption{Time history of the pion propagator at distance 6 for $\kappa =
0.1675$}
\label{fig:timehist}
\end{figure}

\begin{figure}[htb]
\epsfxsize=\columnwidth
\epsffile{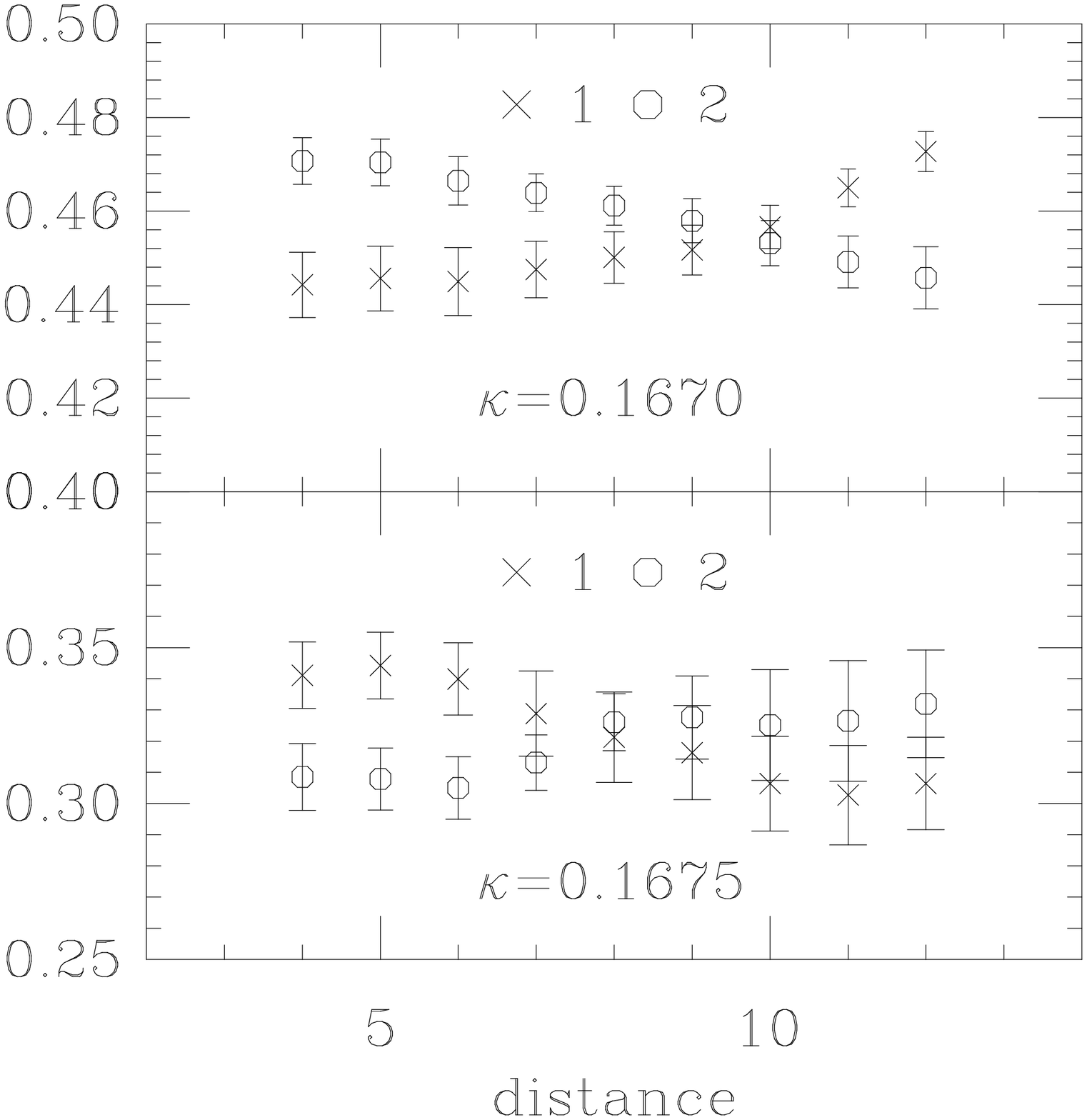}
\caption{Effective pion masses}
\label{fig:eff_mass}
\end{figure}

On every configuration analyzed, we measured from two different wall
sources, one at $t = 0$ (referred to as ``kind" 1) and one at $t = 16$
(``kind" 2). We made fits to the propagators for each ``kind" separately
and also fits to both ``kinds" simultaneously with a common mass but
independent amplitudes. As can be seen from Table~\ref{tab:masses} the two
``kinds" are consistent -- the simultaneous fits ``1 and 2" give a good
confidence level -- except for the pions. We believe the inconsistency
indicates that even the extrapolated errors quoted are still underestimated
due to long autocorrelations in simulation time. A sample time history for
the pion propagator for $\kappa = 0.1675$ is shown in
Fig.~\ref{fig:timehist}. A time history of the pion propagator for $\kappa
= 0.1670$ already appeared in Fig.~4 of reference \cite{doug91}. In
Fig.~\ref{fig:eff_mass} we show the effective mass for the two ``kinds" of
pions at both $\kappa$ values.

\begin{table*}[hbt]
\setlength{\tabcolsep}{1.25pc}
\caption{Results from correlated fits of the propagators to a single
exponential.}
\label{tab:masses}
\begin{tabular}{|ccccccc|}
\hline
 Particle & Kind & $\kappa$ & range & mass & $\chi^2/dof$ &
 C.L. \\
\hline
 pion & 1 & 0.1670 & 11 -- 16 & 0.470( 4) &     20.830/4  &      0.000 \\
 pion & 2 & 0.1670 & 11 -- 16 & 0.450( 4) &      4.483/4  &      0.345 \\
 pion & 1 and 2 & 0.1670 & 10 -- 16 & 0.465( 3) &     52.650/11  &   0.000 \\
 pion & 1  & 0.1675 & 10 -- 16 & 0.322( 9) &      7.396/5  &      0.193 \\
 pion & 2  & 0.1675 & 9 -- 16 & 0.340( 7) &      3.536/6  &      0.739 \\
 pion & 1 and 2 & 0.1675 & 10 -- 16 & 0.312(31) &     14.040/11  &   0.231 \\
\hline
 rho & 1  & 0.1670 & 11 -- 16 & 0.656( 7) &     12.752/4  &      0.013 \\
 rho & 2  & 0.1670 & 9 -- 16 & 0.635( 4) &      5.166/6  &      0.523 \\
 rho & 1 and 2  & 0.1670 & 11 -- 16 & 0.650( 6) &     23.600/9  &   0.005 \\
 rho & 1  & 0.1675 & 5 -- 16 & 0.557( 6) &      6.345/10  &      0.786 \\
 rho & 2  & 0.1675 & 4 -- 16 & 0.531( 5) &     12.790/11  &      0.307 \\
 rho & 1 and 2  & 0.1675 & 6 -- 16 & 0.538( 5) &     11.740/19  &   0.896 \\
\hline
 proton & 1  & 0.1670 & 5 -- 16 & 0.952( 8) &      8.890/10  &      0.543 \\
 proton & 2  & 0.1670 & 5 -- 16 & 0.994( 7) &      4.822/10  &      0.903 \\
 proton & 1 and 2 & 0.1670 & 9 -- 16 & 0.972( 9) &      9.806/13  &   0.710 \\
 proton & 1  & 0.1675 & 5 -- 16 & 0.807(16) &     11.457/10  &      0.323 \\
 proton & 2  & 0.1675 & 4 -- 16 & 0.754( 9) &     14.098/11  &      0.228\\
 proton & 1 and 2 & 0.1675 & 6 -- 16 & 0.804(12) &     16.580/19  &   0.618 \\
\hline
 delta & 1  & 0.1670 & 5 -- 16 & 1.028( 8) &      9.118/10  &      0.521 \\
 delta & 2  & 0.1670 & 5 -- 16 & 1.071(10) &     14.259/10  &      0.161 \\
 delta & 1 and 2 & 0.1670 & 6 -- 16 & 1.061( 7) &     22.710/19  &    0.250 \\
 delta & 1  & 0.1675 & 5 -- 16 & 0.935(10) &     14.464/10  &      0.153 \\
 delta & 2  & 0.1675 & 4 -- 16 & 0.838(13) &      4.361/11  &      0.958 \\
 delta & 1 and 2 & 0.1675 & 7 -- 16 & 0.911(24) &      8.827/17  &    0.946 \\
\hline
\end{tabular}
\end{table*}

\begin{figure}[htb]
\epsfxsize=\columnwidth
\epsffile{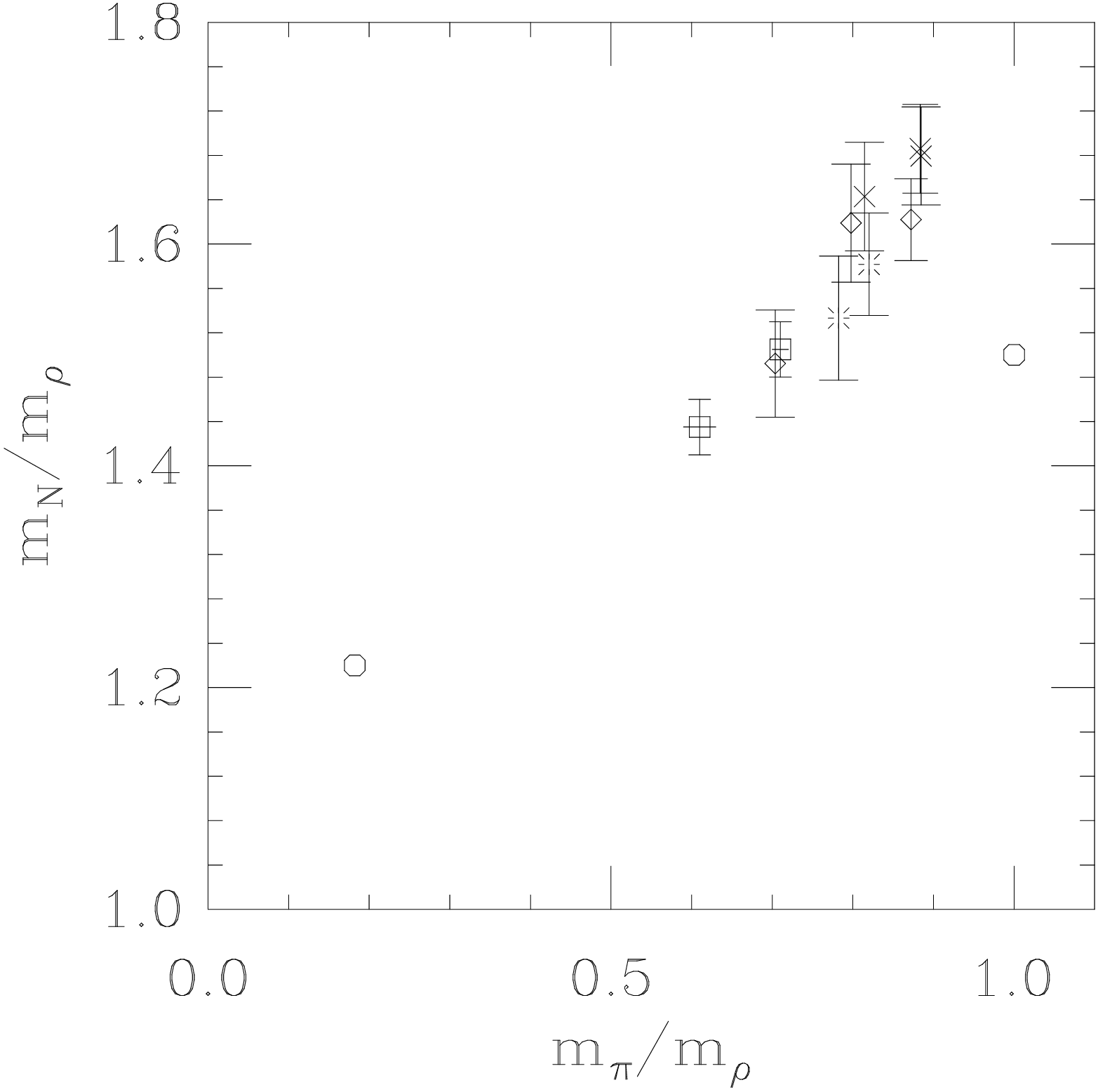}
\caption{Edinburgh plot for dynamical Wilson fermions. Squares are the new
HEMCGC data. All other data come from \protect \cite{rajan}, with crosses
for $\beta = 5.4$, diamonds for $5.5$ and bursts for $5.6$.}
\label{fig:ed}
\end{figure}

With the caveat that our errors might be underestimated in mind, we
computed mass ratios, listed in Table~\ref{tab:ratio}, and show them
--- taking the average over ``kind" 1 and ``kind" 2 ---
together with older dynamical Wilson data from Ref.~\cite{rajan} in the
Edinburgh plot of Fig.~\ref{fig:ed}.
\begin{table}[hbt]
\setlength{\tabcolsep}{0.88pc}
\caption{Mass ratios. The masses and time ranges are the ones from the
previous Table.}
\label{tab:ratio}
\begin{tabular}{|cccc|}
\hline
 Particles & Kind & $\kappa$ & Ratio \\
\hline
 $m_\pi/m_\rho$ & 1 & 0.1670  & 0.71(1)   \\
 $m_\pi/m_\rho$ & 2 & 0.1670  & 0.71(1)   \\
 $m_\pi/m_\rho$ & 1 and 2 & 0.1670  & 0.72(1)   \\
 $m_\pi/m_\rho$ & 1 & 0.1675  & 0.58(2)   \\
 $m_\pi/m_\rho$ & 2 & 0.1675  & 0.64(2)   \\
 $m_\pi/m_\rho$ & 1 and 2 & 0.1675  & 0.58(6)   \\
\hline
 $m_P/m_\rho$ & 1 & 0.1670  & 1.45(2)   \\
 $m_P/m_\rho$ & 2 & 0.1670  & 1.56(3)   \\
 $m_P/m_\rho$ & 1 and 2 & 0.1670  & 1.49(2)   \\
 $m_P/m_\rho$ & 1 & 0.1675  & 1.45(3)   \\
 $m_P/m_\rho$ & 2 & 0.1675  & 1.42(2)   \\
 $m_P/m_\rho$ & 1 and 2 & 0.1675  & 1.49(3)   \\
\hline
\end{tabular}
\end{table}

Extrapolating the pion squared masses (using the average over ``kind" 1 and
``kind" 2) to zero we find a critical kappa value $\kappa_c = 0.16804(6)$.
Extrapolating the other particles to $\kappa_c$ we then obtain $m_\rho =
0.435(12)$, $m_P = 0.57(3)$, and $m_\Delta = 0.71(3)$, giving inverse
lattice spacings of about $1765$, $1645$, and $1735$ MeV, respectively.

In conclusion, we have brought dynamical Wilson fermion spectroscopy almost
up to the level of the staggered one, finding a lightest pion of $m_\pi
\simeq 0.33$, as compared to about $0.27$ for staggered fermions, and $m_\rho
\simeq 0.54$, compared to about $0.52$. While these two flavor simulations
are difficult, {\it i.e.,} CPU time intensive, they come, contrary to the
case of staggered fermions, from an exact algorithm. Our results show that
light pion masses can be achieved with Wilson fermions on finite lattices.
Very preliminary results from a run at $\kappa = 0.1677$ indicate that even
lighter pions ($m_\pi \approx 0.25$ or less) are feasible on present day
lattices.

\section*{Acknowledgements}

This research was supported by the U.~S.~Department of Energy and the
National Science Foundation. The computations were carried out at SCRI,
which is partially funded by DOE.



\begin{thebibliography}{9}
\bibitem{doug91}
D.~Toussaint, {\sl Nucl. Phys. B (Proc Suppl)} {\bf 26}, 3 (1992).
\bibitem{light}
M.~Fukugita, S.~Ohta and A.~Ukawa, {\sl Phys. Rev. Lett.} {\bf 57},
1974 (1986);
Y.~Iwasaki, K.~Kanaya, S.~Sakai and T.~Yoshie, {\sl Phys. Rev. Lett.}
{\bf 69}, 21 (1992).
\bibitem{GC92}
K.~M.~Bitar et al., in preparation.
\bibitem{cmis}
C.~Liu, {\sl Nucl. Phys. B (Proc Suppl)} {\bf 20}, 149 (1991);
A.~D.~Kennedy, {\sl Intl. J. Mod. Phys.} {\bf C3}, 1 (1992).
\bibitem{hemcgc90}
K.~M.~Bitar et al., {\sl Phys. Rev.} {\bf D42}, 3794 (1990).
\bibitem{slgt}
K.~M.~Bitar, R.~G.~Edwards, U.~M.~Heller and A.~D.~Kennedy, these
proceedings.
\bibitem{rajan}
R.~Gupta et al., {\sl Phys. Rev.} {\bf D44}, 3272 (1991).
\end{thebibliography}
\end{document}